\documentclass[preprintnumbers,amsmath,amssymb,nofootinbib,14pt]{revtex4}

\usepackage{hyperref}
\usepackage[utf8]{inputenc}
\usepackage{graphicx}
\usepackage{dcolumn}
\usepackage{bm}
\usepackage{xcolor}
\usepackage{lipsum}
\usepackage{amsmath}
\usepackage{slashed}

\usepackage[a4paper, total={7in, 10.in}]{geometry}

\usepackage{mathtools}

\begin{document}


\title{Thermal History of the Early Universe and Primordial Gravitational Waves from Induced Scalar Perturbations}

\author{Fazlollah Hajkarim}
\email{hajkarim@th.physik.uni-frankfurt.de}

\author{J{\"u}rgen Schaffner-Bielich}
\email{schaffner@astro.physik.uni-frankfurt.de}
  
\affiliation{ Institut f\"ur Theoretische Physik, Goethe Universit\"at, Max von
  Laue Stra{\ss}e 1, D-60438 Frankfurt am Main, Germany }

\date{\today}

\begin{abstract}
We study the induced primordial gravitational waves (GW) coming from the effect of 
scalar perturbation on the tensor perturbation at the second order of cosmological perturbation 
theory. We use the evolution of the standard model 
degrees of freedom with respect to temperature in the early Universe to compute the
induced gravitational waves bakcground. Our result shows that the spectrum of the induced 
GW is affected differently by the standard model degrees of freedom than the GW coming
from first order tensor perturbation. This phenomenon is due to the presence of scalar perturbations as a source for tensor perturbations and  it is effective around the quark gluon deconfinement and electroweak transition. In case of considering a scalar spectral index larger than one at small scales or a non-Gaussian curvature power spectrum this effect can be observed by gravitational wave observatories.
\end{abstract}

\maketitle

\section{\label{sec:intro}Introduction}

The first observation of gravitational wave (GW) event from binary black holes merger  and also binary neutron stars 
by LIGO opened a new way to our understanding about the Universe \cite{Abbott:2016nmj,Abbott:2016blz,TheLIGOScientific:2017qsa}.
Before using electromagnetic waves we could see the physical phenomena in the 
Universe, however now somehow we can listen to them in case 
they produce gravity waves. Beyond GW produced by astrophysical events at the 
recent era of cosmic evolution, it would be interesting to listen to the early moments 
of the Universe to detect the predicted GW by cosmological and particle physics models \cite{Caprini:2018mtu,Maggiore:1999vm}. 
The inflationary scenario as a theory to explain the moments after the big bang 
and explain the current cosmological observations can produce 
 that kind of primordial gravitational waves (PGW) \cite{Grishchuk:1974ny, Starobinsky:1979ty}. 
 In principle if we have detectors of high sensitivity, one can observe such PGW. 
However, their detection depends on the first order tensor to scalar  ratio ``r"  which appears  directly
in the predicted observable relic density of tensor perturbation at the first order of cosmological perturbation
theory. Since the cosmic microwave background (CMB) experiments like Planck have only put an upper bound on 
the tensor-to-scalar ratio and some theoretical predictions assume very tiny values close to zero for this quantity, it motivates to consider the second order terms in the cosmological perturbation theory 
induced by scalar perturbations. 
This has been done in the literature \cite{Baumann:2007zm,Assadullahi:2009nf,Ananda:2006af,Mollerach:2003nq}, and  gained much interest recently 
\cite {Kohri:2018awv,Inomata:2018epa,Drees:2019xpp,Inomata:2019zqy,Inomata:2019ivs,Lu:2019sti,Cai:2018dig,Cai:2019elf,Unal:2018yaa,Cai:2019amo,Ben-Dayan:2019gll,Gong:2019mui,Tomikawa:2019tvi,Espinosa:2018eve} due to the first direct discovery of GW signals from binary mergers.

The predicted shape for the induced spectrum using the standard model (SM) equation
of state at different scales or frequency and the evolution of induced tensor power spectrum should be roughly like the first order one. However, this has not been checked numerically and the goal of this paper is to do exactly this.
Studying induced PGW is also interesting when there is non-Gaussianity
 on scales smaller than the cosmic microwave background (CMB) \cite{Unal:2018yaa,Chen:2010xka}. This can
  boost the spectrum of induced PGW to large values of the relic density which might be 
  accessible by near future experiments \cite{Cai:2018dig,Cai:2019elf,Unal:2018yaa} especially by pulsar timing arrays. 
Also, it would be one of the real chances to test particle physics models and early 
Universe cosmology based on our current data from CMB
and standard model of particle physics and $\Lambda$CDM cosmology in the case that 
 tensor-to-scalar perturbation ratio is much smaller than the current bound from Planck \cite{Aghanim:2018eyx} or from future CMB experiments like CMBS4 \cite{Abazajian:2019eic}.

  Currently, GW experiments can mostly detect the GW spectrum from astrophysical events with a strong enough amplitude. However, future 
  detectors will be able to probe much smaller signatures from possible phase transitions and the PGW in the early Universe cosmology. Some of these 
  planned GW missions are the 
Laser Interferometer Space Antenna (LISA)~\cite{Audley:2017drz},  the Einstein Telescope (ET)~\cite{Sathyaprakash:2012jk}, the Deci-hertz Interferometer Gravitational wave Observatory (DECIGO and B-DECIGO)~\cite{Seto:2001qf, Sato:2017dkf}, the Big Bang Observatory  
(BBO)~\cite{Crowder:2005nr},  
 the Square Kilometer Array (SKA) telescope ~\cite{Janssen:2014dka}, the North American Nanohertz Observatory for Gravitational Waves (NanoGrav) \cite{Arzoumanian:2018saf}, and the European Pulsar Timing Array (EPTA) \cite{Lentati:2015qwp}. Moreover, 
LIGO can be sensitive to the predicted GW for the early Universe depending on the model \cite{LIGOScientific:2019vic,Abbott:2017xzg}.

For a pre-BBN cosmology which is dominated by some matter with an equation of state different from radiation the spectrum of
the first order and the induced PGW will be different as studied in refs.~\cite{Inomata:2018epa,Kohri:2018awv,Inomata:2019zqy,Inomata:2019ivs,Assadullahi:2009nf,Bernal:2019lpc,Figueroa:2019paj,Opferkuch:2019zbd,Alabidi:2013wtp}. Such models will not be the focus of the current paper.

The value of secondary produced PGW depends on the amplitude of scalar perturbations which is observed at the CMB scale \cite{Aghanim:2018eyx}. Since the scalar spectral index at large scales is highly constrained by CMB and is close to one, it is not expected to deviate significantly from a flat scalar power spectrum for the standard case. However, it might be different at smaller scales e.g. at pre-BBN era.
This makes the induced PGW different from the first order one which has different 
 free parameters i.e. the scalar-to-tensor ratio and the tensor spectral index \cite{Saikawa:2018rcs,Bernal:2019lpc,Watanabe:2006qe}.

In the next section (\ref{sec:sm}) we investigate the thermal history of the early Universe including SM particles.
 In Sec. \ref{sec:perturb}  the set of equations related to first order and second order 
 perturbations and the induced PGW will be discussed. We compute the relic density of the induced 
 PGW for scale invariant and scale dependent scalar perturbations in Sec. \ref{sec:pgw}. 
 Finally, we summarize our results in Sec.~\ref{sec:conclusion}.

\section{Thermal History of the Early Universe and Standard Model Particles}
\label{sec:sm}
\begin{figure}
\includegraphics[scale=0.9]{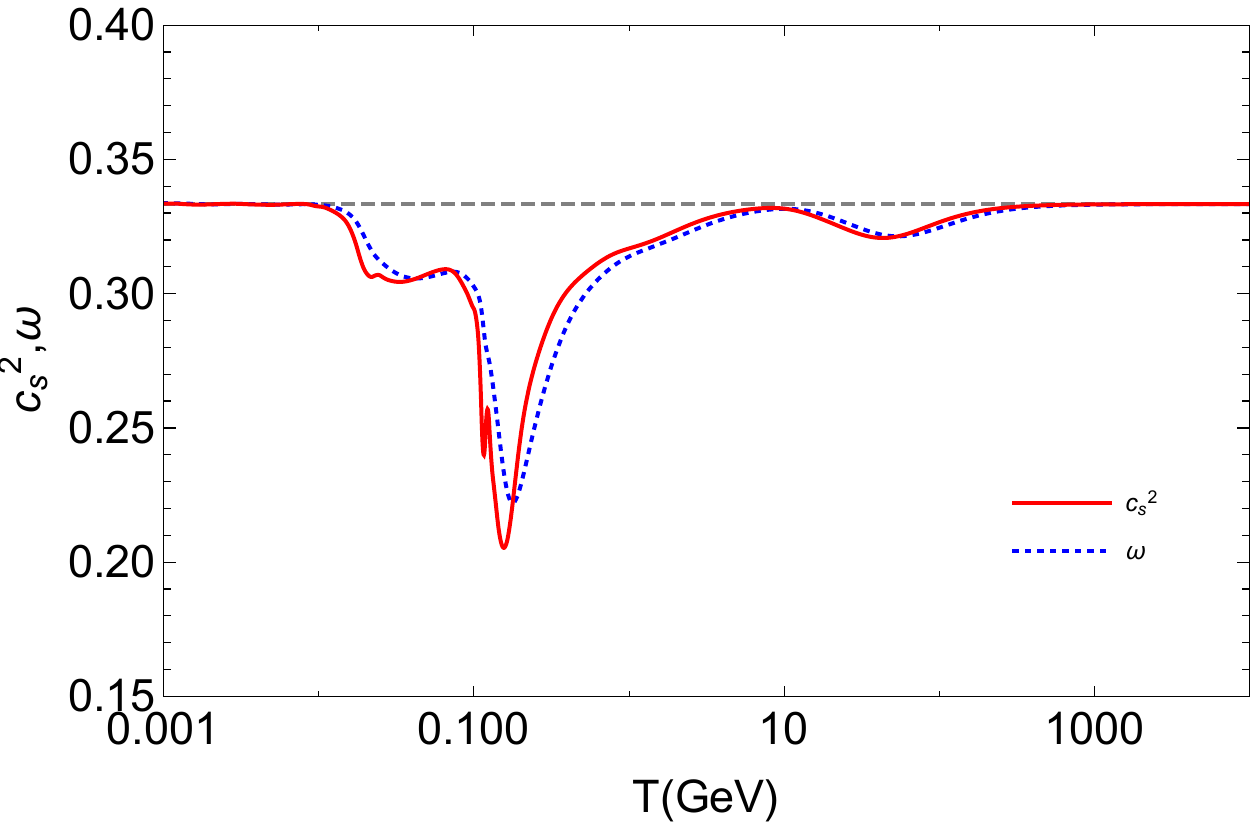}
\caption{\label{fig:soundspeedo} The speed of sound $c_s^2$ (solid red curve) and equation of state parameter $\omega$ (blue dotted curve) with respect to temperature $T$ including all SM degrees of freedom at relevant temperatures using the result of ref.~\cite{Drees:2015exa}. The dashed line shows the ideal gas case $c_s^2=\omega=1/3$. }
\end{figure}

 The effect of standard model degrees of freedom on the (first order) PGW especially 
 around the quark gluon deconfinement and electroweak transition are studied in the literature \cite{Schettler:2010dp,Hajkarim:2019csy,Saikawa:2018rcs,Bernal:2019lpc,Watanabe:2006qe}.
The evolution of degrees of freedom influences the relation between the scale factor and 
the temperature in the early Universe cosmology \cite{Hajkarim:2019csy,Drees:2015exa}. However, this can be interpreted as an 
evolution of the equation of state parameter. The equation of state parameter and 
speed of sound can be driven by other thermodynamics variables as 
 \begin{eqnarray}
\omega=\frac{p_{\text{tot}}}{\rho_{\text{tot}}}\,,\,\,\,\,\,c_s^2=\frac{\partial p_{\text{tot}}}{\partial \rho_{\text{tot}}}\,,\,\,\,\,\, p_{\text{tot}}= Ts_{\text{tot}}-\rho_{\text{tot}}\,,
\end{eqnarray}
where the energy density and entropy density are defined as
 \begin{eqnarray}
  \rho_{\text{tot}}=\frac{\pi^2}{30}g_{\text{eff}}(T)T^4\,,\,\,\,\,\,
s_{\text{tot}}=\frac{2\pi^2}{45}h_{\text{eff}}(T)T^3\,,
\end{eqnarray}
and the effective relativistic degrees of freedom (DoF) for the energy and the entropy density are denoted by $g_{\text{eff}}$ and $h_{\text{eff}}$.
The contribution of every relativistic fermion (boson) to the effective DoF is $7/8$ ($1$).
For radiation or any ideal relativistic fluid $c_s^2=\omega=1/3$.
We use the result of ref.~\cite{Drees:2015exa} to compute
 the equation of state parameter and speed of sound for SM degrees of freedom as plotted in fig.~\ref{fig:soundspeedo}.
This figure shows 
 that around temperatures of $150$~MeV and $100$~GeV, due to quantum chromodynamics (QCD) and electroweak crossover transitions $\omega$ and $c_s^2$ will be smaller than $1/3$. Due to the interaction between particles 
 the thermal bath of the early Universe deviates from an ideal relativistic fluid. 
 Every extra degree of freedom e.g. due to dark matter or any beyond the SM physics 
in the early Universe can also lead to small effects on these quantities. Especially, 
considering a highly populated sector like supersymmetry, can modify the result of  fig.~\ref{fig:soundspeedo} and add a new  valley at temperatures around $1$~TeV or higher depending on the scale of supersymmetry (see refs. \cite{Jinno:2013xqa,Cui:2018rwi,Watanabe:2006qe} for supersymmetric DoF and PGW).
 
\section{Tensor  and  Scalar Perturbation Equations}
\label{sec:perturb}

 At second order of the cosmic perturbation theory one can consider the following metric \cite{Baumann:2007zm,Ananda:2006af,Kohri:2018awv}
 \begin{eqnarray}
 \label{metnew}
\text{d}s^2 = g_{\mu\nu}\text{d}x^\mu \text{d}x^\nu = -a^2(1+2\Phi) \text{d}\eta^2 + a^2 \left( (1-2\Psi)\delta_{ij}+\frac{1}{2}h_{ij} \right) \text{d}x^i \text{d}x^j,
\end{eqnarray}
where $a$ is the scale factor and $\eta$ the conformal time.  Assuming the metric does not contain anisotropic stress ($\Phi=\Psi$) the scalar perturbation and tensor perturbation are denoted by $\Phi$  and $h$, respectively. The metric in eq.~(\ref{metnew}) is considered in Newtonian gauge. Here we mainly follow ref.~\cite{Kohri:2018awv}  for the evolution equations of induced scalar and tensor perturbations. 
The evolution of the tensor power spectrum sourced from scalar perturbation can 
be obtained by solving the following equation in Fourier space \cite{Baumann:2007zm,Assadullahi:2009nf,Ananda:2006af,Kohri:2018awv}
\begin{eqnarray}
\label{tensorpert}
h''_{\vec{ k}}(\eta) + 2 \mathcal{H} h'_{\vec{ k}}(\eta)+ k^2 h_{\vec{ k}}(\eta) =& 4 S_{\vec{ k}}(\eta),  \label{EOM_h}
\end{eqnarray}
where the conformal Hubble parameter is denoted by $\mathcal{H}=aH$ and the derivative with respect to the conformal time by $^\prime=d/d\eta$. The Hubble parameter can be evaluated from the Friedmann equation $H^2=(8\pi/3M_{Pl}^2)\rho_{tot}$, with  $M_{Pl}$ being the Planck mass. Also, the entropy conservation in the early Universe is considered\footnote{
To find the evolution of temperature with respect to scale factor in the standard cosmology after reheating the entropy conservation should be assumed. This can be taken into account by the relation 
\begin{eqnarray}
\frac{dT}{da}=-\frac{\frac{T}{a}}{1+\frac{T}{3h_{\text{eff}}(T)}\frac{dh_{\text{eff}}(T)}{dT}}\,.
\end{eqnarray}}. To consider the whole evolution of DoF for finding $a(\eta)$ one should solve $a^\prime=a^2H$ numerically. 

The source 
term on the right hand side of eq.~(\ref{tensorpert}) assuming $\Phi=\Psi$ can be evaluated as 
\cite{Baumann:2007zm,Assadullahi:2009nf,Ananda:2006af,Kohri:2018awv}
\begin{eqnarray}
\label{sourceterm}
S_{\vec{ k}} =& \int \frac{\text{d}^3 l}{(2 \pi)^{3/2}} \epsilon_{ij}({\vec{ k}}) l_i l_j \left( 2\Phi_{\vec{ l}}  \Phi_{{\vec{ k}}-{\vec{ l}}} + \frac{4}{3(1+\omega)} \left( \mathcal{H}^{-1} \Phi'_{\vec{ l}} + \Phi_{\vec{ l}}\right) \left( \mathcal{H}^{-1} \Phi'_{{\vec{ k}}-{\vec{ l}}} + \Phi_{{\vec{ k}}-{\vec{ l}}} \right)  \right)\,,
\end{eqnarray}
where the polarization tensor for GW is shown by $\epsilon_{ij}({\vec{ k}})$. 
In eq.~(\ref{sourceterm}) the terms depending on $c_s^2$ cancel out so that the expression depends only on $\omega$ explicitly.
We use the Green's function method to solve the differential equation for tensor perturbations. 
  Then the solution in the comoving frame will be \cite{Kohri:2018awv}
\begin{eqnarray}
a(\eta) h_{\vec{ k}}(\eta) = 4 \int^\eta_{\eta_{\text{i}}} \text{d}\tilde{\eta} \mathcal{G}_{\vec{ k}}(\eta, \tilde{\eta}) a(\tilde{\eta}) S_{\vec{ k}}(\tilde{\eta}),
\end{eqnarray}
where we make use of the Green's function as the solution of the following equation \cite{Baumann:2007zm,Kohri:2018awv}
\begin{eqnarray}
\label{greenfun}
\mathcal{G}_{\vec{ k}}''(\eta, \tilde{\eta}) +\left( k^2 - \frac{ a''(\eta)}{a(\eta)}\right) \mathcal{G}_{\vec{ k}}(\eta, \tilde{\eta}) = \delta (\eta - \tilde{\eta}). \label{EOM_Green}
\end{eqnarray}
For a fixed $\omega$ the scale factor evolves as $a\propto \eta^{\frac{2}{3\omega+1}}$. Then  eq.~(\ref{greenfun}) will be given by 
\begin{eqnarray}
\label{greenfunom}
\mathcal{G}_{\vec{ k}}''(\eta, \tilde{\eta}) +\left(k^2 - \left[\frac{2}{3\omega+1}\right]\left[\frac{2}{3\omega+1}-1\right]\frac{1}{\eta^2}\right) \mathcal{G}_{\vec{ k}}(\eta, \tilde{\eta}) = \delta (\eta - \tilde{\eta}). \label{EOM_Green}
\end{eqnarray}
To compute the Green's function we use the method of ref.~\cite{Baumann:2007zm}.
 If the functions $\mathcal{G}_1$ and $\mathcal{G}_2$ are two homogeneous solutions of eq.~(\ref{greenfun}) for a mode $k$ then
the Green's function will be defined as 
\begin{eqnarray}
\label{greennum}
\mathcal{G}(\eta,\tilde{\eta})=\frac{\mathcal{G}_1(\eta)\mathcal{G}_2(\tilde{\eta})-\mathcal{G}_1(\tilde{\eta})\mathcal{G}_2(\eta)}{\mathcal{G}_1^\prime(\tilde{\eta})\mathcal{G}_2(\tilde{\eta})-\mathcal{G}_1(\tilde{\eta})\mathcal{G}_2^\prime(\tilde{\eta})}\,,
\end{eqnarray}
where in practice we compute the numerical form of this function (assuming $\mathcal{G}_1(\eta)=0$, $\mathcal{G}_1^\prime(\eta)=1$; $\mathcal{G}_2(\eta)=1$, $\mathcal{G}_2^\prime(\eta)=0$) for each mode $k$ or GW frequency $f=2\pi/k$.

The evolution of scalar perturbations in cosmology can be described with the differential equation \cite{Kohri:2018awv,Mukhanov:2005sc}
\begin{eqnarray}
\label{scalperful}
\Phi''_{\vec{ k}} + 3 \mathcal{H} (1 + c_{\text{s}}^2) \Phi'_{\vec{ k}} + (2 \mathcal{H}'+(1+3 c_{\text{s}}^2)\mathcal{H}^2 +c_{\text{s}}^2 k^2) \Phi_{\vec{ k}}  = \frac{a^2}{2} \tau \delta \mathcal{S},  \label{EOM_Phi_complete}
\end{eqnarray}
where $\delta p=c_s^2\delta \rho+\tau \delta \mathcal{S}$. By assuming $c_s^2 = \omega=const.$ and 
vanishing entropy perturbation in cosmology $\delta \mathcal{S}=0$ one gets \cite{Kohri:2018awv}
\begin{eqnarray}
\label{scalpertlim}
\Phi''_{\vec{ k}}(\eta) + \frac{6(1+\omega)}{(1+3\omega)\eta } \Phi'_{\vec{ k}}(\eta) + \omega k^2 \Phi_{\vec{ k}}(\eta)=0. \label{EOM_Phi}
\end{eqnarray}
In the limit $c_s^2 \rightarrow \omega$ our numerical results match with the 
analytical solution of eq.~(\ref{scalpertlim}) as given in refs.~\cite{Kohri:2018awv,Baumann:2007zm}.  
However, for a precise numerical calculation one should distinguish between the speed of sound $c_s^2$
 and the equation of state parameter $\omega$ as shown in fig.~\ref{fig:scalargw} and consider 
them separately in the computation.
The correlation function for curvature perturbations is derived as shown in ref. \cite{Kohri:2018awv} and gives 
\begin{eqnarray}
\label{scalarpow}
\langle \phi_{\vec{ k}} \phi_{ \vec{\bar{k}}} \rangle = \delta ({\vec{ k}}+{\vec{\bar{ k}})} \frac{2\pi^2}{k^3} \left( \frac{3[1+\omega]}{5+3\omega} \right)^2 \mathcal{P}_\mathcal{R} (k)\,,
\end{eqnarray}
where the superhorizon value of $\phi_{{\vec{ k}}}$ is evaluated from the relation $\Phi_{\vec{ k}}=\Phi(k\eta)\phi_{{\vec{ k}}}$.
The transfer function $\Phi(k\eta)$ reaches one before the moment of horizon crossing ($k=a_{hc}H_{hc}$). 
We will consider the evolution of the degrees of freedom of SM in the result 
of eq.~(\ref{scalperful}) and other relevant equations in the following. 

The induced tensor power spectrum from scalar perturbations is determined via \cite{Baumann:2007zm,Ananda:2006af,Kohri:2018awv}
\begin{eqnarray}
\label{powtenint}
\mathcal{P}_T^{{}} (\eta, k) =  4  
 \int_0^\infty \int_{\left| 1-v \right |}^{1+v}\text{d}v ~\text{d} u \left[ \frac{4v^2 - (1+v^2-u^2)^2}{4vu} \right]^2 \mathcal{I}^2 (v,u,x) \mathcal{P}_\mathcal{R} ( k v ) \mathcal{P}_\mathcal{R} ( k u )\,, \label{PT}
\end{eqnarray}
where $v=|\vec{l}|/|\vec{k}|$ and $u=|\vec{k}-\vec{l}|/|\vec{k}|$. 
In eq.~(\ref{PT}) the definition of the function $\mathcal{I}$ reads \cite{Ananda:2006af,Kohri:2018awv}
\begin{eqnarray}
\label{eq:I}
\mathcal{I}(v,u,x)= \int_0^x \text{d}\tilde{x} \frac{a (\tilde{\eta})}{a(\eta)} k \mathcal{G}_k (\eta, \tilde{\eta}) f (v,u,\tilde{x}),   \label{I}
\end{eqnarray}
where the source function $f$ based on the equation of state parameter $\omega$ and transfer functions $\Phi$ is obtained from \cite{Ananda:2006af,Kohri:2018awv}
\begin{eqnarray}
\label{eq:f}
f (v ,u ,\tilde{x}) = & \frac{6(\omega+1)}{3\omega+5}\Phi(v\tilde{x})\Phi(u\tilde{x})+\frac{12(\omega+1)}{(3\omega+5)^2\mathcal{H}} \left[ \partial_{\tilde{\eta}}\Phi(v\tilde{x})\Phi(u\tilde{x})+\partial_{\tilde{\eta}} \Phi(u\tilde{x})\Phi(v\tilde{x}) \right]
+\frac{12(1+\omega)}{(3\omega+5)^2\mathcal{H}^2}  \partial_{\tilde{\eta}}\Phi(v\tilde{x})\partial_{\tilde{\eta}}\Phi(u\tilde{x})\,.
\end{eqnarray}
The terms depending on $c_s^2$ for the case $\Phi=\Psi$ in the original equation of 
refs.~\cite{ Baumann:2007zm,Ananda:2006af} cancel out. Then only the terms on the right hand side of eq.~(\ref{eq:f}) are left.
By changing the variables $u$ and $v$ to  $u=(p+q+1)/2$ and $v=(p-q+1)/2$ \cite{Kohri:2018awv} the numerical solution of the power spectrum turns out to be computationally faster.

For different values of $\omega$ shown in fig.~\ref{fig:tenspow} we derive the Green's function and $\Phi$  from eqs.~(\ref{greennum}) and (\ref{scalpertlim}) and computed then the function $f$ from eq.~(\ref{eq:f}). Then, the function $\mathcal{I}$ is computed analytically from eq.~(\ref{eq:I}) and used in eq.~(\ref{PT}) for numerical integration. 
As an example, the Green's function for $\omega=1/3$ considering $a\propto\eta$ and $a^{\prime\prime}=0$ from eq.~(\ref{greenfunom}) can be evaluated as $\mathcal{G}_k(\eta,\tilde{\eta})=\sin(\eta-\tilde{\eta})/k$ \cite{Kohri:2018awv}. Also, the scalar transfer function for a radiation like fluid by solving eq.~(\ref{scalpertlim}) and assuming ($\Phi(k\eta\rightarrow 0)\rightarrow 1$) can be shown to be  \cite{Kohri:2018awv,Ando:2018qdb}
\begin{eqnarray}
\Phi(k\eta)=\frac{9}{(k\eta)^2}\left[\frac{\sin (k\eta/\sqrt{3})}{k\eta/\sqrt{3}}-\cos (k\eta/\sqrt{3})\right]\,.
\end{eqnarray}
The explicit form of the function $\mathcal{I}$ is complicated and long and will not be presented here (see refs.~\cite{Ananda:2006af,Kohri:2018awv}). 
In ref.~\cite{Kohri:2018awv} the limits of the function $\mathcal{I}$ in a radiation dominated Universe are found to be $\mathcal{I}(u,v,x\rightarrow 0)\approx x^2/2$  and $\mathcal{I}(u,v,x\rightarrow\infty)\approx \mathcal{I}(u,v)/x^2$. Using the latter limit the evolution of the tensor power spectrum is estimated to be $\mathcal{P}(\eta,k)/A_{\mathcal{R}}^2\approx 19.7/x^2$ \cite{Kohri:2018awv}. This expression is vastly used afterwards to estimate the value of the GW relic density of today. However, the value $19.7$ does not match the results for $\omega=1/3$ shown by the solid black line in fig.~\ref{fig:tenspow} based on a complete numerical calculation. This will be explained in more details in the next section.

 \section{Primordial Gravitational Waves and Experimental Constraints}
 \label{sec:pgw}
 \begin{figure}
\includegraphics[scale=0.64]{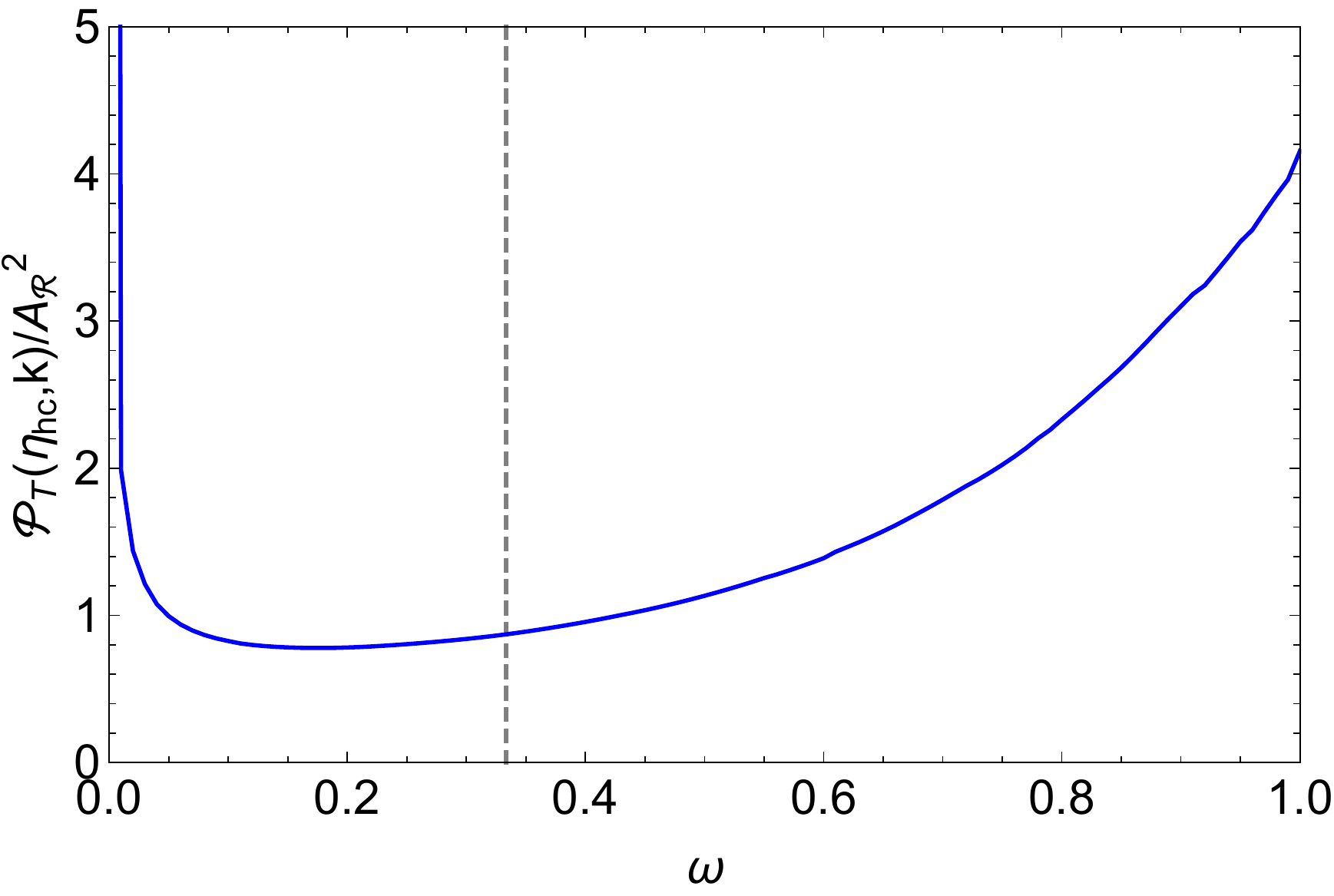}
\caption{\label{fig:omtenper} Evolution of the scaled tensor perturbation at horizon crossing with respect to a constant equation of state parameter $\omega$ of the background fluid (blue solid curve). The vertical grey dashed line shows the ideal gas case $\omega=1/3$.}
\end{figure}

\begin{figure}
\includegraphics[scale=0.84]{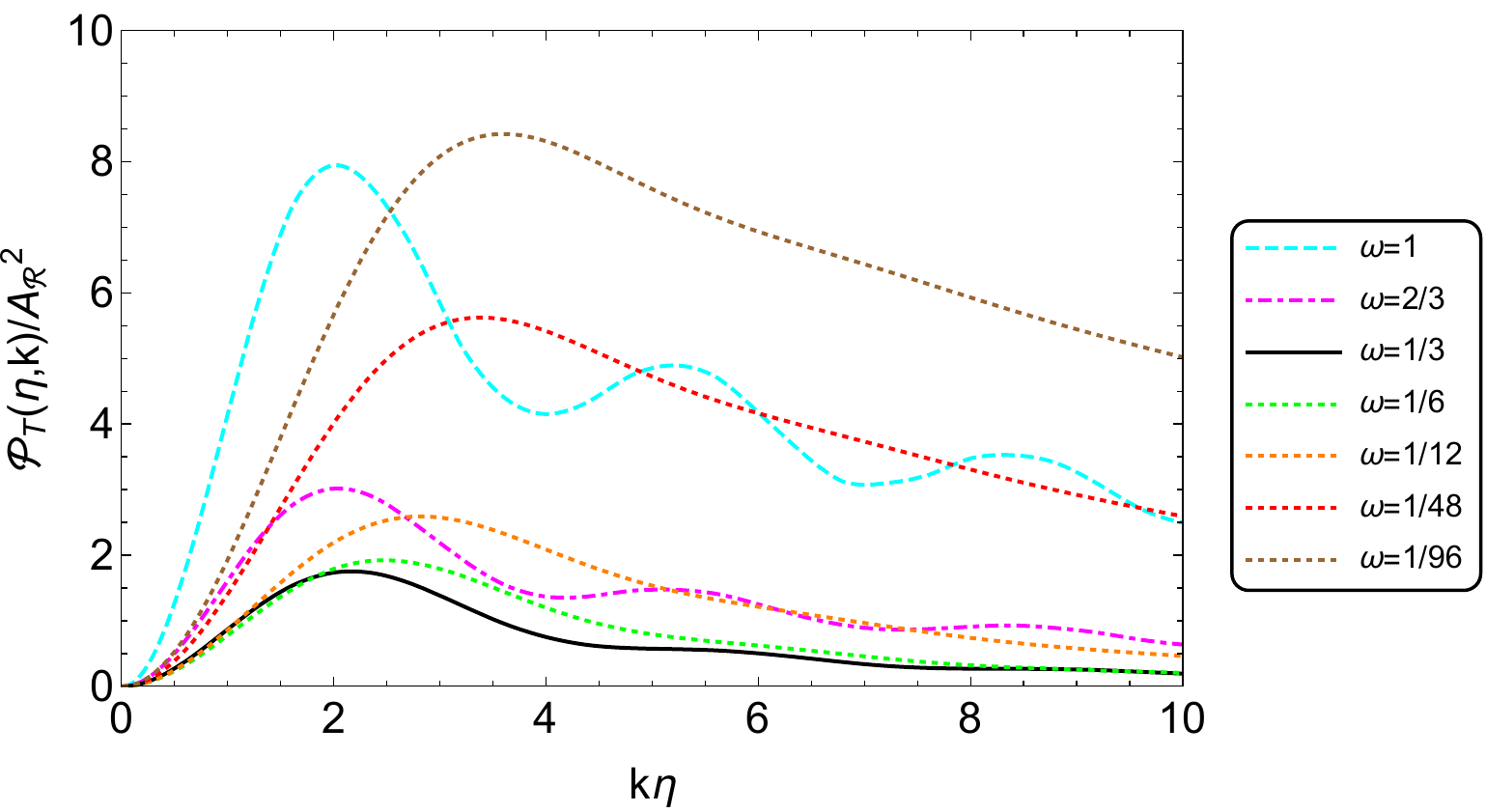}
\caption{\label{fig:tenspow} Evolution of the scaled tensor power spectrum $\mathcal{P}(\eta,k)/A_{\mathcal{R}}^2$ with
 respect to $k\eta$ for different values of $\omega$ is shown. As $\omega\rightarrow 0$ the tensor power spectrum diverges as the upper limit of first integral in eq.~(\ref{powtenint}) becomes infinite \cite{Kohri:2018awv}.
 }
\end{figure}
 
 The scale invariant scalar power spectrum defined in eq.~(\ref{scalarpow}) can be written as \cite{Baumann:2009ds}
 \begin{eqnarray}
 \mathcal{P}_{\mathcal{R}}(k)=A_{\mathcal{R}}\,. 
 \end{eqnarray}
Moreover, the power law scalar power spectrum can be defined as \cite{Baumann:2009ds}
 \begin{eqnarray}
 \mathcal{P}_{\mathcal{R}}(k)=A_{\mathcal{R}} \left(\frac{k}{\tilde{k}}\right)^{n_s-1}\,,
 \end{eqnarray}
with the pivot scale $\tilde{k}=0.05$~Mpc$^{-1}$ and the scalar spectral index $n_s$.

 From Planck satellite observations the value of the scalar spectral index has been determined to be  $n_s=0.965\pm 0.004$ \cite{Aghanim:2018eyx}. Also, the value of the curvature perturbation amplitude at the CMB scale is $A_{\mathcal{R}}\simeq2.1\times10^{-9}$ \cite{Aghanim:2018eyx}.
In addition, the scale invariant tensor power spectrum at first order is defined as \cite{Baumann:2009ds}
\begin{eqnarray}
\mathcal{P}_T^{(1)}=A_T\,.
\end{eqnarray}
 Then the tensor to scalar ratio at first order is \cite{Baumann:2009ds} 
\begin{eqnarray}
r=\frac{A_T}{A_{\mathcal{R}}}\,. 
\end{eqnarray}
The current limit from the Planck collaboration is $r\lesssim0.07$ \cite{Aghanim:2018eyx}.

 In fig.~\ref{fig:omtenper} we assume that the parameter $\omega$ is fixed and constant in the calculation of 
 eqs.~(\ref{greenfun}), (\ref{scalpertlim}), and (\ref{powtenint}). The scaled tensor power spectrum (scaled by the square of the scale invariant curvature power spectrum: $\mathcal{P}_T^{{}}(\eta_{hc},k)/A_{\mathcal{R}}^2$) at 
 horizon crossing is computed for the range of $0<\omega<1$ where the minimum of the tensor power spectrum appears at $\omega_{\text{min}}=0.175$. As fig.~\ref{fig:omtenper}  shows and as it is analytically calculated in refs.~\cite{Kohri:2018awv,Assadullahi:2009nf,Alabidi:2013wtp} if $\omega \rightarrow0$ the tensor power spectrum diverges $\mathcal{P}_T^{{}}(\eta \geq \eta_{hc},k)\rightarrow \infty$. The final result for $\mathcal{P}_T^{{}}(\eta_{hc},k)/A_{\mathcal{R}}^2$ in the range $0\leq k\eta\leq10$ and different $\omega$'s is shown in fig.~\ref{fig:tenspow}. 
 The function $\mathcal{I}$ inside the integral of eq.~(\ref{PT}) is highly oscillatory. It gets
  its maximum value slightly after  horizon crossing ($2\lesssim k\eta \lesssim 4$ depending 
  on the value of $\omega$) then decays, as $\propto 1/(k\eta)^2$ during radiation domination \cite{Kohri:2018awv}. 
 
  For the induced tensor perturbation in the computation of the tensor power spectrum after
 horizon crossing until today the WKB method can not be used to match the result at some point and 
 then extend it until today, since  (to the best of our knowledge) there is not any well defined shape of the transfer function during radiation domination for the induced PGW 
 like $A \sin (\delta + k \eta)/k\eta$ as assumed in refs.~\cite{Saikawa:2018rcs,Watanabe:2006qe,Bernal:2019lpc} for the first order PGW.

 The power spectrum, shown in fig.~\ref{fig:tenspow}, increases after horizon 
crossing to a maximum value then dilutes with the expansion of the Universe. The numerical 
calculation of such a power spectrum is 
difficult and time consuming since it includes various numerical integrations over oscillatory functions. In practice it is very difficult to solve for a large range of $k\eta$ from zero to today $k\eta_0$. So one should calculate until horizon crossing or until the peak for $\omega=1/3$ in fig.~\ref{fig:tenspow} then, use 
$1/(k\eta)^2$ as the dilution factor due to the expansion of the Universe.

 The tensor power spectrum can also be defined in terms of the transfer function $\mathcal{P}(\eta_{hc},k)\propto \mathcal{T}^\prime(\eta_{hc},k)^2=k^2\mathcal{T}(\eta_{hc},k)^2$. It scales like $1/a^2$ after horizon crossing. For the radiation domination case (without the change of DoF) the scaled tensor power spectrum in fig.~\ref{fig:tenspow} at horizon crossing and at its first peak ($\tilde{x}_{\text{1st,peak}}\simeq2.17$) are $\mathcal{P}_T(\eta_{hc},k)/A_{\mathcal{R}}^2\simeq 0.87$ and $\mathcal{P}_T(\eta_{\text{1st,peak}},k)/A_{\mathcal{R}}^2\simeq 1.75$, respectively ($\mathcal{P}_T(\eta_{\text{1st,peak}},k)/\mathcal{P}_T(\eta_{hc},k)\simeq2$).  
 
\begin{figure}
\includegraphics[scale=0.65]{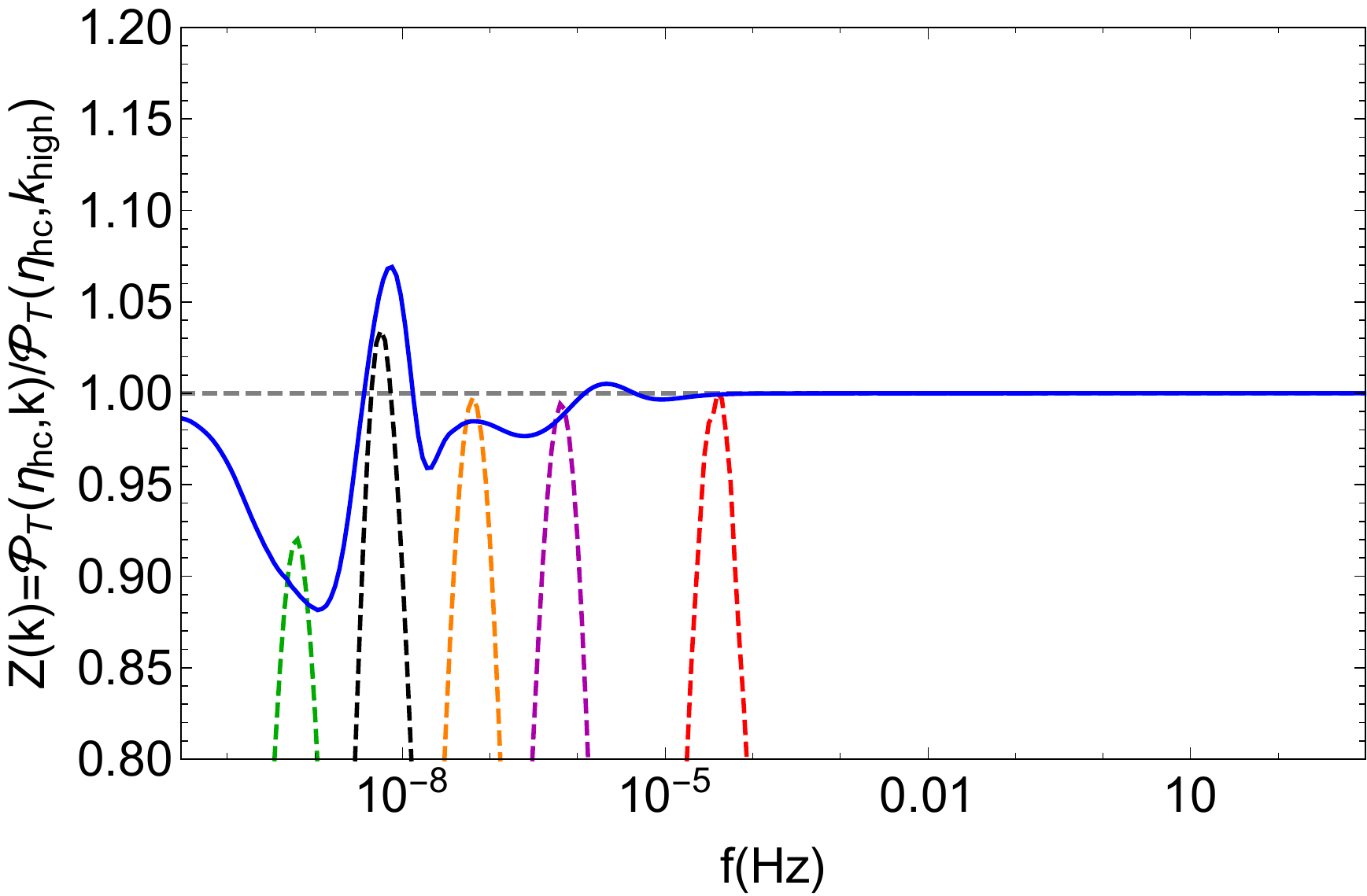}
\caption{\label{fig:scalargw} The correction factor $Z(k)=\mathcal{P}_T(\eta_{hc},k)/\mathcal{P}_T(\eta_{hc},k_{\text{high}})$ on the scalar induced gravitational
 waves due to the evolution of standard model degrees of freedom which is scaled to the value at high frequencies where $\omega\rightarrow 1/3$ well above the electroweak transition. The solid blue line includes the 
 thermal history using all SM particles. The dashed grey straight line denotes a
 radiation like equation of state with $\omega=1/3$.  The correction factor for non-Gaussianity peaks around frequencies $\bar{k}/2\pi=\bar{f}=1.6\times10^{-9}$ Hz (green), $1.6\times10^{-8}$ Hz  (black), $1.6\times10^{-7}$ Hz  (orange), $1.6\times10^{-6}$ Hz (violet), and $10^{-4}$ Hz  (red) is plotted in colored dashed curves for the choice of $[\mathcal{F}_{NL}^2, A_G]= [10 ,10^{-2}]$.  }
\end{figure}

It is expected that the evolution of the DoF in the early Universe plays a role in the computation of  eqs.~(\ref{greenfun}),  (\ref{scalperful}) and (\ref{PT}) so we checked it. We have included all DoF of the SM to find the scaled tensor power spectrum $\mathcal{P}_T(\eta_{hc},k)/A_{\mathcal{R}}^2$. Then the shape of the spectrum will have a correction factor, shown in fig.~\ref{fig:scalargw}, due to the evolution of different functions inside the integral of eq.~(\ref{PT}) by a change of the DoF over time. This fact  originates from the combination of all functions in the integral $a(\eta)/a(\tilde{\eta})$, $f(u,v,k\tilde{\eta})$, and $\mathcal{G}(\eta,\tilde{\eta})$ due to carrying the retarded effects from previous times. In practice, an analytical formula to show the explicit form of fig.~\ref{fig:scalargw} has not been found. However, it roughly behaves like the evolution of  $d\omega/dT$ including the retardation effects due to the Green's function. 
The inflection points in the correction factor of secondary PGW spectrum roughly show the points of the crossover transition having the smallest values of $c_s^2$ and $\omega$ at the relevant frequencies ($3\times10^{-9}$~Hz and $10^{-6}$~Hz). In the case that one adds the supersymmetric DoF or any highly populated sector around the TeV scale, there will be another inflection point at higher frequencies around $10^{-4}$~Hz.

\begin{figure}
\includegraphics[scale=0.5]{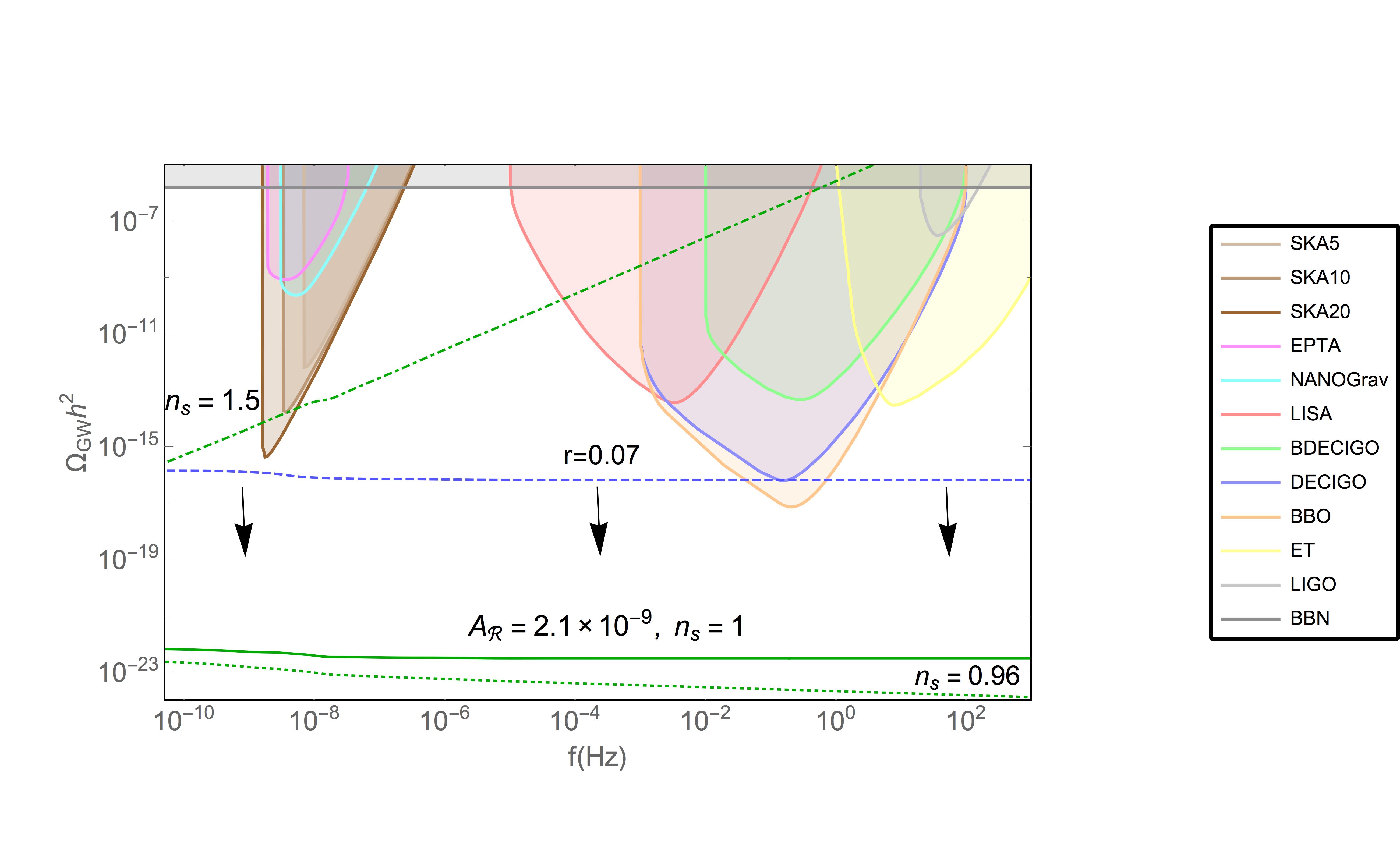}
\caption{\label{fig:pgwexp} The spectrum of the PGW from the scale invariant first order tensor perturbations (blue dashed line)
assuming the maximum value of tensor-to-scalar ratio $r=0.07$ from the upper limit of  Planck data \cite{Aghanim:2018eyx} and the induced PGW 
using the observed value of scalar perturbation amplitude $A_\mathcal{R}=2.1\times10^{-9}$ assuming the scale invariance (green solid line), a scale dependence with $n_s=0.96$ (green dotted line), and $n_s=1.5$ (green dot-dashed line)
as shown in the plot. Also, some current and future experimental constraints \cite{Audley:2017drz,Sathyaprakash:2012jk,Seto:2001qf, Sato:2017dkf,Crowder:2005nr,Janssen:2014dka,Arzoumanian:2018saf,Lentati:2015qwp,LIGOScientific:2019vic,Abbott:2017xzg} and the BBN bound are 
shown (see the text for details).}
\end{figure}

The density of PGW per frequency (wave number) from tensor perturbation at first order in the case of eq.~(\ref{tensorpert}) is source free and can be estimated as \cite{Saikawa:2018rcs,Bernal:2019lpc,Watanabe:2006qe}
 \begin{eqnarray}
  \label{gwrelic1st}
\Omega_{GW}^{(1)}(\eta_0,k)h^2=\frac{1}{24} \left(\frac{a_{hc}H_{hc}}{a_0H_0}\right)^2 \langle{\mathcal{P}_T^{(1)}(\eta_0,k)}\rangle_{\text{osc}}\simeq \frac{1}{24}\left(\frac{h_{\text{eff}}(T_0)}{h_{\text{eff}}(T_{hc})}\right)^{\frac43}\left(\frac{g_{\text{eff}}(T_{hc})}{g_{\text{eff}}(T_0)}\right)\mathcal{P}_T^{(1)}(k)\Omega_{\gamma,0}h^2 \,,
\end{eqnarray}
 where ${\mathcal{P}_T}^{(1)}(k)=\mathcal{P}_T ^{(1)}(\eta,k)/{\mathcal{T}^\prime}^{(1)}(\eta,k)^2=\mathcal{P}_T^{(1)}(\eta_{hc},k)/(k {\mathcal{T}}^{(1)}(\eta_{hc},k))^2$ for 
 the first order PGW when ${\mathcal{T}}^{(1)}(\eta_{hc},k)^2$ is the $1$st order tensor perturbation  transfer function.  Here we focus on frequencies larger than $3\times10^{-11}$~Hz which are equivalent to temperatures larger than $4$~MeV. Consequently, we do not consider the effects of neutrinos and photons free streaming which appear at smaller frequencies \cite{Weinberg:2003ur,Stefanek:2012hj,Saikawa:2018rcs} as a new source term on the right hand side of eq.~(\ref{tensorpert}).
 The relic PGW at today from induced scalar perturbation 
 can be obtained from \cite{Ando:2018qdb}
 \begin{eqnarray}
 \label{gwrelic2nd}
\Omega_{GW}^{{}}(\eta_0,k)h^2=\frac{1}{24} \left(\frac{a_{hc}H_{hc}}{a_0H_0}\right)^2 \langle{\mathcal{P}_T^{{}}(\eta_0,k)}\rangle_{\text{osc}}\simeq  
\frac{1}{24}\left(\frac{h_{\text{eff}}(T_0)}{h_{\text{eff}}(T_{hc})}\right)^{\frac43}\left(\frac{g_{\text{eff}}(T_{hc})}{g_{\text{eff}}(T_0)}\right) 2\times (2.17)^2 Z(k) {\mathcal{P}_T^{{}}(\eta_{hc},k_{\text{high}})} \Omega_{\gamma,0}h^2\,,
\end{eqnarray}
the correction factor is defined as $Z(k)=\mathcal{P}_T^{{}}(\eta_{hc},k)/\mathcal{P}_T^{{}}(\eta_{hc},k_{\text{high}})$ shown in fig.~\ref{fig:scalargw} as blue solid line ($\mathcal{P}_T^{{}}(\eta_{hc},k_{\text{high}})/A_{\mathcal{R}}^2\simeq 0.87$). 
The shape of this function completely depends on the details of the functions inside the integral of eq.~(\ref{powtenint}). The function $Z(k)$ already corrects the spectrum on the order of $10\%$ especially around the QCD epoch which is equivalent to a frequency of $3\times10^{-9}$~Hz. 
As mentioned earlier the evolution of DoF of SM particles is taken from the results of ref.~\cite{Drees:2015exa}. Using the numerical factor $2\times(2.17)^2$ in eq.~(\ref{gwrelic2nd}) from the fact that $\tilde{x}_{\text{1st,peak}}\simeq2.17$, the tensor power spectrum from scale invariant curvature power spectrum during radiation domination evolves as $[\mathcal{P}_T(\eta_{\text{1st,peak}},k)\tilde{x}_{\text{1st,peak}}^2]/[\mathcal{P}_T(\eta_{hc},k)\tilde{x}_{hc}^2]\times 1/\tilde{x}^2\simeq 2\times(2.17)^2 /\tilde{x}^2$.
The results for the first order and second order PGW are shown in fig.~\ref{fig:pgwexp} in addition to the different experimental constraints \cite{Audley:2017drz,Sathyaprakash:2012jk,Seto:2001qf, Sato:2017dkf,Crowder:2005nr,Janssen:2014dka,Arzoumanian:2018saf,Lentati:2015qwp,LIGOScientific:2019vic,Abbott:2017xzg} and the BBN bound \cite{Kohri:2018awv} as outlined in the legend of the plot (the suffix number for the SKA experiment shows the constraints based on the number of years in operation).
In case of scale invariance the relic of the first order PGW can have any value from $\sim10^{-16}$ down to the lower limit by the induced PGW of $\sim10^{-23}$ depending on the value of $r$. If $r\lesssim 10^{-9}$ then the scalar induced contribution on the PGW will be dominating. The scale dependent induced PGW for scalar spectral indices of $n_s=0.96$ and $1.5$ are also plotted in fig.~\ref{fig:pgwexp}. For the case $n_s>1$ the spectrum can be constrained by current and future experiments.

\begin{figure}
\includegraphics[scale=0.5]{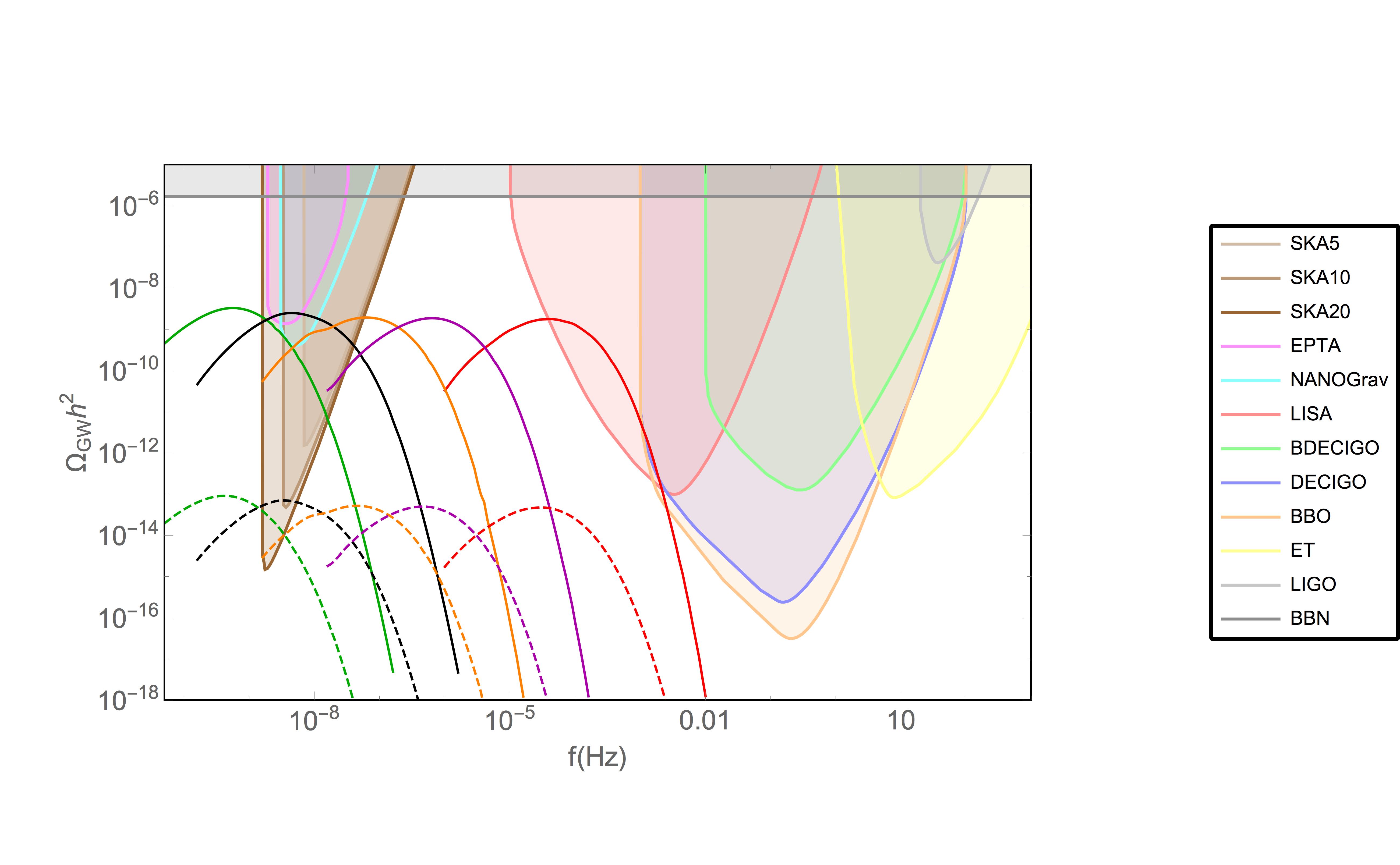}
\caption{\label{fig:pgwexpnongauss} The induced PGW from non-Gaussian scalar power spectrum at small scales is shown for characteristic frequencies $\bar{k}/2\pi=\bar{f}=1.6\times10^{-9}$ Hz (green), $1.6\times10^{-8}$ Hz  (black), $1.6\times10^{-7}$ Hz  (orange), $1.6\times10^{-6}$ Hz (violet), and $10^{-4}$ Hz  (red) for two choices of $[\mathcal{F}_{NL}^2, A_G]= [10 ,10^{-2}]$ (solid curves) and $ [60 ,10^{-4}]$ (dashed curves). }
\end{figure}

The curvature power spectrum can deviate from a flat shape at scales smaller than the CMB scale especially if one assumes that there is some Gaussian or non-Gaussian source for it from the curvature bispectrum \cite{Cai:2018dig,Unal:2018yaa,Chen:2010xka}. 
This can enhance the spectrum by some peaks of the relic densities  at frequencies accessible by pulsar timing arrays or other experiments. The effect of the evolution of the degrees of freedom in the early Universe also appears on the secondary PGW if there is some source of non-Gaussianity \cite{Cai:2018dig,Unal:2018yaa}.  To represent this effect we assume first that the following Gaussian curvature power spectrum (for $k$ in the logarithmic scale) around a specific mode $\bar{k}$ with width $\bar{\sigma}$ \cite{Cai:2018dig,Cai:2019elf,Unal:2018yaa} is parameterised by 
\begin{eqnarray}
\label{powGauss}
\mathcal{P}_{\mathcal{R},G}=A_G\exp\left[{-\frac{\ln(k/\bar{k})^2}{2\bar{\sigma}^2}}\right]\,,
\end{eqnarray}
where $A_G$ is the Gaussian amplitude that can depend on the characteristic scale $\bar{k}$ and other model dependent variables. Here we simply assume it to be constant to test for its observational consequence.
By considering the presence of non-Gaussianity in the curvature power spectrum one gets 
\cite{Cai:2018dig,Cai:2019elf,Unal:2018yaa}
\begin{eqnarray}
\label{pownon-Gauss}
\mathcal{P}_{\mathcal{R},NG}(k)=\mathcal{P}_{\mathcal{R},G}(k)+\mathcal{F}_{NL}^2\int^\infty_0\int^{1+v}_{|1-v|} \text{d}v ~\text{d} u \frac{\mathcal{P}_{\mathcal{R},G}(kv)\mathcal{P}_{\mathcal{R},G}(ku)}{v^2u^2}\,, 
\end{eqnarray}
where the nonlinear factor for the non-Gaussian curvature perturbation is denoted by $\mathcal{F}_{NL}$. 
We choose different benchmark points for $\bar{k}$ and $\mathcal{F}_{NL}$ to see how  its consequence will be on the PGW. Also, we consider the evolution of DoF to check how it influences the induced PGW by non-Gaussian power spectrum.
 This is also shown in fig.~\ref{fig:scalargw} for the scaled tensor power spectrum assuming $\mathcal{F}_{NL}^2=10$, $A_G=10^{-2}$, and $\bar{\sigma}=1$ for different $\bar{k}$'s. As we expect from previous discussion the effect of QCD and electroweak transitions is also apparent in this case for different peaks around modes $\bar{k}/2\pi=\bar{f}=1.6\times10^{-9}$, $1.6\times10^{-8}$, $1.6\times10^{-7}$, and $1.6\times10^{-6}$ Hz. The value of scaled tensor power spectrum in for the mentioned modes is scaled to its value at a high frequency $\bar{f}=10^{-4}$ Hz.
We should emphasize again that this effect is due to the changes of the equation of state parameter and DoF during cosmic transitions (see fig.~\ref{fig:soundspeedo}) in the calculation of eq.~(\ref{powtenint}).
  
Non-Gaussianity at the CMB scale is constrained by Planck data \cite{Akrami:2019izv}. However, there are  less constraints on non-Gaussianity at small scales. To predict a realistic signal for the presence of non-Gaussianity we assume $A_G\mathcal{F}_{NL}^2\ll1$ \cite{Unal:2018yaa}.  This causes that the second term on the right hand side of eq.~(\ref{pownon-Gauss}) is smaller than the first term which means that the non-Gaussian part of the curvature power spectrum is smaller than the Gaussian one. However, the contribution of non-Gaussianity on the second order PGW peak can be larger than the Gaussian part depending on the values of $\mathcal{F}_{NL}$ and $\bar{\sigma}$ \cite{Unal:2018yaa}.  In fig.~\ref{fig:pgwexpnongauss} the non-Gaussian induced PGW for $[\mathcal{F}_{NL}^2, A_G]= [10 ,10^{-2}]$ and  $[60 ,10^{-4}]$ for different modes is plotted using eqs.~(\ref{gwrelic2nd}), (\ref{powGauss}), and (\ref{pownon-Gauss}). For these peaks pulsar timing array experiments or LISA can be sensitive enough to observe such peaks on the GW background which can be distinguished from other prediction, e.g. by first order phase transitions, due to the specific shape of the spectrum. Consequently, the effect of the
thermal history of the Universe and the induced tensor spectra from scalar perturbations will be small at the peaks and can provide information about pre-BBN cosmology.

\section{Discussion and Conclusions} 
\label{sec:conclusion}
The second order PGW from scalar perturbations have 
been studied analytically and by a numerical calculation
 in the radiation and early matter dominated epochs \cite{Baumann:2007zm,Assadullahi:2009nf,Ananda:2006af,Mollerach:2003nq,Inomata:2018epa,Kohri:2018awv,Drees:2019xpp,Inomata:2019zqy,Inomata:2019ivs,Lu:2019sti,Cai:2018dig,Cai:2019elf,Unal:2018yaa,Cai:2019amo,Ben-Dayan:2019gll}.  Here first we focused on studying the effect of the equation of state parameter on the induced PGW during the dominance of radiation or any fluid with $0<\omega<1$ (see fig.~\ref{fig:omtenper}). 
 The shape of the tensor spectrum is shown in fig.~\ref{fig:tenspow} for different equation of state parameters for $0\leq k\eta\leq10$. 
Moreover, the effect of the degrees of freedom of thermal bath particles evolving with temperature on 
 the spectrum of induced PGWs at different frequencies is studied and shown by a correction function $Z(k)$ ($Z(T_{hc})$) in eq.~(\ref{gwrelic2nd}) and fig.~\ref{fig:scalargw}. This function includes the evolution of the tensor perturbation and the retarded effects from the scalar Green's function and is particularly important around the QCD and electroweak transitions. It is possible that future experiments do not 
 observe the first order PGW for a scale independent tensor spectral index and tiny values of the tensor to scalar ratio $\sim10^{-9}$
 at first order, based on the current knowledge of $\Lambda$CDM cosmology and 
  observations. Then it is expected that the induced PGW should be observable when
 GW experiments can probe the PGW relic down to $\sim5\times10^{-23}$ (fig.~\ref{fig:pgwexp}). Larger values due to non-Gaussianity (fig.~\ref{fig:pgwexpnongauss}) are measurable by the recently planned missions,  
 and are distinguishable from astrophysical backgrounds. In a sense 
 what is calculated here is the combined prediction of general relativity based on just the SM of 
 particle physics using the current CMB observation for the early Universe cosmology. 

Principally, one should be able to observe the impact of thermal history of the early Universe on the
induced PGW background from scalar 
perturbations. Since we have already observed the scalar perturbation at the CMB scale
and their effect on structure formation, it would be reasonable to see 
its indirect impact on the production of induced PGW at smaller scales. Otherwise, there should be some theoretical model dependent explanation for it from alternative approaches to the big bang cosmology and inflation or an unknown effect at early epochs that has not been investigated before \cite{Kohri:2018awv,Baumann:2007zm,Ananda:2006af,Ben-Dayan:2019gll}.

Considering the effect of the standard model DoF on the induced PGW will be influential for the 
future searches of gravitational waves with cosmological origin, since it helps to improve the theoretical prediction due to different effects for the GW spectra from the early Universe. In addition, studying details of the induced PGW affects  
our understanding about the impact of thermal evolution of the SM in the 
early Universe and pre-BBN cosmology on the GW background.
Also, searching for the PGW can be useful as a cosmic laboratory 
for the indirect probe of new particles being present in the thermal bath
 of the early Universe, and new physics even without 
observing any boosted effects on the GW spectrum from the predicted cosmic
 phase transitions and nonstandard cosmological scenarios.

\begin{acknowledgments}

FH thanks Caner {\"U}nal for useful discussions. 
Also, he is grateful to the hospitality and support of
 Galileo Galilei Institute for Theoretical Physics at Florence during the last stages of this work. 
The authors acknowledge the support
  by the Deutsche Forschungsgemeinschaft (DFG)
  through the CRC-TR 211 
  project number 315477589-TRR 211.
  
\end{acknowledgments}


\bibliography{inducedgw.bib}

\end{document}